\documentclass{article} 
\usepackage{psfig, amssymb}

\def\baselinestretch{1.5}
\begin{document}
\newcommand{\gen}[1]{\mbox{$\langle #1 \rangle$}}
\newcommand{\mchoose}[2]{\def\baselinestretch{1} \left ( {
 {#1} \choose{#2} } \right ) \def\baselinestretch{2}}

\def\Box{[]}
\newenvironment{proof}{\noindent{\sc Proof:\ }}{\vspace{2ex}}
\newenvironment{prooft}{{\bf

Proof of the Theorem.}}{$\Box$ \vspace{1ex}}
\newenvironment{prooff}{{\bf

Proof.}}{\vspace{1ex}}
\newenvironment{proofl}{{\bf

Proof.}}{$\Box$ \vspace{1ex}}

%Eqn numbers within sections and theorems, lemmas ... within sections
% remember to put \setcounter{equation}{0} at start of every section

\def\numberlikeadb{\global\def\theequation{\thesection.\arabic{equation}}}
\numberlikeadb
\newtheorem{theorem}{Theorem}[section]
\newtheorem{lemma}[theorem]{Lemma}
\newtheorem{corollary}[theorem]{Corollary}
\newtheorem{proposition}[theorem]{Proposition}
\newtheorem{example}[theorem]{Example}

\newcommand{\hh}{{\hspace{.3cm}}}
\newcommand{\RR}{{\bf R}}
\newcommand{\NN}{{\bf N}}
\newcommand{\ZZ}{{\bf Z}}
\newcommand{\PP}{{\bf P}}
\newcommand{\EE}{{\bf E}}
\newcommand{\Var}{{\mbox{Var}}}
\newcommand{\beas}{\begin{eqnarray*}}
\newcommand{\enas}{\end{eqnarray*}}

\newcommand{\bea}{\begin{eqnarray}}
\newcommand{\ena}{\end{eqnarray}}
\newcommand{\eq}{\begin{equation}}
\newcommand{\en}{\end{equation}}

\def\comm#1{\footnote{#1 }}

\def\gcomm#1{\comm{(Gesine)---#1}}
\def\grcomm#1{\comm{(Gesine)---Could not read this word#1}}
\def\acomm#1{\comm{(Andrew)---#1}}
\def\ignore#1{}

\def\half{{\textstyle{\frac12}}}
\def\be{{\bf e}}
\def\uo{^{(0)}}
\def\ul{^{(l)}}
\def\bM{{\bf M}}
\def\bW{{\bf W}}
\def\Ref#1{(\ref{#1})}
\def\a{\alpha}
\def\w{\omega}
\def\l{\lambda}
\def\pnmw{\PP_{NMW}}
\def\whv{{\widehat V}}
\def\fij{\phi_{ij}}
\def\Xij{X_{ij}}
\def\pij{p_{ij}}
\def\siim{\sum_{i=1}^m}
\def\sjn{\sum_{j=1}^n}
\def\siimx{\sum_{i=1}^{M_x}}
\def\sjnx{\sum_{j=1}^{N_x}}
\def\quarter{{\tfrac14}}
\def\ffti{{\cal F}_t^\infty}
\def\Bi{{\rm Bi}\,}
\def\dtv{d_{TV}}
\def\law{{\cal L}}
\def\Zij{Z_{ij}}
\def\ep{\hfill $\Box$ 

\bigskip}
\def\Tr#1{Theorem~\ref{#1}}
\def\giv{\,|\,}
\def\remark{\noindent {\bf Remark.}\ \,}
\def\Yil{Y_{il}}
\def\tfrac#1#2{{\textstyle{\frac#1#2}}}

\def\Po{{\rm Po\,}}
\def\BHJ{\cite{bhj}}
\def\non{\nonumber}
\def\th{\theta}
\def\bN{{\bf N}}
\def\e{\varepsilon}
\def\m{\mu}
\def\uil{^{(1,l)}}
\def\var{{\rm Var\,}}
\def\dd{\delta}
\def\r{\rho}
\def\t{\tau}
\def\s{\sigma}
\def\z{\zeta}
\def\bs{{\bf s}}
\def\bu{{\bf u}}
\def\ex{\EE}
\def\tH{{\widetilde H}}
\def\Blb{\left\{}
\def\Brb{\right\}}
\def\Bl{\left(}
\def\Br{\right)}
\def\Blm{\left|}
\def\Brm{\right|}

\title{Small worlds}
\author{
A. D. Barbour\\
Applied Mathematics\\
University of Z\"urich\\
CH - 8057 Z\"urich\\
and
\\
Gesine Reinert\\
King's College Research Centre\\
UK - Cambridge CB2 1ST}

\maketitle

\begin{abstract}
Small world models are networks consisting of many local links
and fewer long range `shortcuts'.  In this paper, we consider
some particular instances, and rigorously investigate the
distribution of their inter--point network distances.
Our results are framed in terms of approximations, whose
accuracy increases with the size of the network.  We also
give some insight into how the reduction in typical
inter--point distances occasioned by the presence of shortcuts
is related to the dimension of the underlying space.
\end{abstract}

\section{Introduction}
 \setcounter{equation}{0}

In \cite{WattsStrogatz}, Watts and Strogatz introduced a mathematical
model for ``small--world'' networks. These networks had achieved
popularity in social sciences, modelling the phenomenon of ``six degrees
of separation''. 
Further examples that have been suggested are the neural network of
C.elegans, the power grid of the western United States, and the
collaboration graph of film actors. 
 The work of  \cite{WattsStrogatz} has received considerable attention
during the last two years, in particular by physicists; see the Los
Alamos server for condensed matter physics ({\tt
http://xxx.lanl.gov/archive/cond-mat}). 
However, a closely related model, the ``great circle model'', had
already
 been studied by Ball {\it et al.\/} \cite{Balletal}, in the context of
epidemics. 

The model  proposed in  \cite{WattsStrogatz} is as follows. Starting
from a 
ring lattice (a 1-dimensional finite lattice with periodic boundary
conditions) with $L$ vertices, the~$k$ nearest neighbors to a vertex 
in clockwise direction are connected to the vertex by an undirected
edge, 
resulting in a  $2k$--nearest neighbour graph. Next, each edge is
rewired at random with probability~$p$. The procedure for this is: a
vertex is chosen, and an edge that connects it to its nearest neighbour
in a clockwise sense. With probability~$p$, this edge is reconnected to
a vertex chosen uniformly at random over the entire ring, with duplicate
edges forbidden; otherwise the edge is left in place. The process is
repeated by moving clockwise around the ring, considering each vertex in
turn until one lap is completed. Next, the edges that connect vertices
to their second-nearest neighbours are considered, as before. So the
rewiring process stops after~$k$ laps. The quantities computed from this
graph are the ``characteristic path length'' $\ell(p)$, defined as the
number of edges in the shortest path between two vertices, averaged over
all pairs of vertices, and the ``clustering coefficient'' $C(p)$,
defined as follows. Suppose that a vertex $v$  has $k_v$ neighbours. Let
$C_v$ be the quotient of the number of edges between these $k_v$
neighbours and the possible number of edges ${k_v \choose 2}$. Define
$C$ to be the average of $C_v$ over all $v$. These two quantities are
computed for real networks in the examples above, as well as for the
random (Bernoulli) graph with the same $p$. The common phenomenon observed 
is the ``small-world phenomenon'': that $\ell$ is not much larger
than $\ell_{random}$, but that $C \gg
C_{random}$. 

Ideally, from a probabilistic view point, one would like to determine
the behaviour of $\ell(p)$ and $C(p)$ so as to estimate the
parameter~$p$ in the network, or so as to be able to assign statistical
significance levels, when distinguishing different network models.
Physicists seem to be intrigued by the scaling properties of $\ell$; see
\cite{NewmanWatts}, \cite{NewmanWatts2}, and references therein; also
they enjoy studying percolation on this graph; see  \cite{MooreNewman}, 
\cite{NewmanWatts}. Percolation there is also viewed as a model for
disease spread.  

A closer look reveals that the above model is not easy to analyze. In
particular, there is a nonzero probability of having isolated vertices,
which makes~$\ell(p)$ infinite with positive probability, and hence $\EE
\ell(p)
= \infty$. As a result, it
 was soon revised by not rewiring edges, but rather adding edges, thus
ensuring that the graph stays connected; see, for example, 
\cite{NewmanWatts}. More precisely, a number of shortcuts are added
between randomly chosen pairs of sites with probability $\phi$ per
connection on the underlying lattice, of which there are $Lk$.  Thus, on
average, there are $Lk\phi$ shortcuts in the graph. 

Recently, Newman, Moore  and Watts \cite{NewmanWatts},
\cite{NewmanWatts2} 
gave a heuristic computation 
(the NMW heuristic) of $\ell$ in this modified graph. They suggest
 that
\bea \label{nw}
\EE \ell = \frac{L}{k} f(Lk\phi),
\ena 
where
\beas
f(z)=\frac{1}{2 \sqrt{z^2 + 2 z}} \tanh^{-1} \sqrt{\frac{z}{z+2}}.
\enas
In particular,
\bea \label{nwa}
f(z) \sim \left\{ \begin{array}{ll}
                     \frac{1}{4} & \mbox{ for } z \ll 1\\
                     \frac{\log 2z}{4z}  & \mbox{ for } z \gg 1.
                         \end{array}
                          \right. ,
\ena 
giving 
\beas
\EE \ell \sim \left\{ \begin{array}{ll}
                     \frac{L}{4k} & \mbox{ for } Lk\phi \ll 1\\
                     \frac{\log 2Lk\phi}{4k^2 \phi}  & \mbox{ for }
Lk\phi \gg 1.
                         \end{array}
                          \right. .
\enas
Their heuristic is based on  mean
field approximations,  replacing random variables by their 
expectations.

  In the context of epidemics in a spatially structured population,  
 Ball {\it et al.} consider individuals on a large circle in their 
``great circle model''. They allow only nearest--neighbour ($k=1$)
connections, but claim that most of their results can be extended to
general~$k$.
An SIR epidemic is studied, where each individual has a probability
$p_L$ of infecting a neighbour, and probability $p_G$ of
infecting any other 
individual on the circle; typically, $p_L \gg p_G$. In the SIR
framework, this model corresponds to individuals having a fixed
infectious period of duration~$1$.  The structure of the graph at time
$T= \infty$ is their main
object of interest. In terms of small worlds, their model broadly 
corresponds to having an epidemic on a small-world network with 
parameter~$\phi$, where $\phi = p_G/p_L$. 

In this paper, we analyze a continuous model, introduced 
in~\cite{MooreNewman}, in which a random number of chords, with
Poisson distribution $\Po(L\rho/2)$, are uniformly and independently
superimposed as shortcuts on a circle of circumference~$L$. Distance is
measured as usual along the circumference, and chords are deemed to be
of length zero.  When~$L$ is large, this model approximates the 
$k$--neighbour model 
of~\cite{NewmanWatts} if $\rho = 2k\phi$, except that distances should
also be divided by~$k$, because unit graph distance in the
$k$--neighbour
model covers an arc length of~$k$, rather than~$1$.  In the case when
the expected number $L\rho/2$ of shortcuts is large, we prove a
distributional approximation for the distance between a randomly chosen
pair of points $P$ and~$P'$, and give a bound on the order of the error, 
in terms of total variation distance.  This distribution differs, in
both location and spread, from that suggested by the NMW heuristic,
though to the coarsest order $O(\frac{1}{\rho}\log (L\rho))$ agrees with
that
suggested by~\Ref{nwa}.  We also show that analogous results can be
proved in higher dimensions by much the same method, when the circle
is replaced by a sphere or a torus; here, the reduction in the typical
distance between pairs of points occasioned by shortcuts is less
dramatic than in one dimension.

\section{The continuous circle model: construction and heuristics}
 \setcounter{equation}{0} 

In this section, we consider a continuous model consisting of a 
circle $C$ of circumference $L$, to which are added a Poisson 
$\Po(L\rho/2)$ number of 
uniform and independent random chords. We begin with a dynamic
realization of 
 the network, which describes, for each $t\ge0$, the set of points $R(t)
\subset C$ that can be reached from a given point~$P$ within time~$t$:
time corresponds to arc distance, with chords of length zero.  Such a
realization is also the basis for the NMW heuristic.

Pick Poisson $\Po(L\rho)$ points of the circle $C$ uniformly and
independently, 
and call this set
 $Q$. The elements $q \in Q$ are called  {\it potential\/} label~1 end
points of chords. To each $q \in Q$, assign a second independent uniform
point of $C$, say $q'=q'(q)$: the label~2 end point.  The unordered
pairs $\{q,
q'\}$ form the {\it potential\/} chords. 
Only a random subset of the potential chords are actually realized.
Let $R(t)$ be the union of the~$B(t)$ intervals of $C$, each of which 
increases with time, growing deterministically at rate $1$
at each end point; we start with $R(0)=\{P\}$. 
Whenever $card(\partial R(t-) \cap Q) = 1$ --- that is, whenever the 
boundary of $R(t-)$ reaches a
potential label~1 end of a chord (note that
the intersection never contains more than one element, with
probability~$1$)
--- so that 
$\partial R(t-) \cap Q = \{q\}$, say, accept the chord $\{q, q' \}$ if
$q'
\not\in R(t-)$ (that is, if the chord would reach beyond the cluster
$R(t-)$) 
and take $R(t) = {\overline{R(t-)}}\cup \{q'\}$; otherwise, take $R(t) =
{\overline{R(t-)}}$. This defines a
predictable thinning of the set of potential chords, to obtain the set
of actual, accepted chords.  The number of intervals increases by~$1$
whenever a
chord is accepted, and decreases by~$1$ whenever two intervals grow into
one another.  

The
intensity of adding chords is $2 \rho B(t) \{1 - r(t) L^{-1}\} $, 
with the label~2 end points uniformly distributed over $R^c(t)$, where
$r(t) = |R(t)|$, the Lebesgue measure of $R(t)$. The integrated
intensity 
is thus
\beas
\int_0^\infty 2 \rho B(t) \{1 - r(t) L^{-1}\} dt = L\rho /2 \hh \mbox{
a.s.}, 
\enas
since $\frac{d}{dt} r(t) = 2 B(t)$ a.e.\ with respect to
 Lebesgue measure, and 
 $$
r(\infty) = r(L/2) = L.
$$ 
 Thus we generate $\Po(L\rho/2)$ accepted chords. 
To see that they are uniformly chosen, 
simply note that this is true of the potential chords, and that each
potential chord is accepted with probability~$1/2$, independently of the
number and positions of all potential chords, according to whether the
first of its end points to belong to~$R$ had initially been chosen as
the label~1 or the label~2 end point. Thus this growth and merge
construction 
indeed results in $\Po(L\rho/2)$ chords, uniformly 
distributed over $C$. 

The NMW heuristic takes the equation
\bea \label{nmw1}
\frac{d}{dt} r(t) = 2 B(t),
\ena
and adds to it an equation
\bea \label{nmw2}
\frac {d}{dt} B(t) = 2\rho B(t) \{1-r(t)L^{-1}\} - 2
B(t)(B(t)-1)(L-r(t))^{-1},
\ena
derived by treating the discrete variable $B(t)$ as continuous and the
corresponding jump rates as differential rates.  The final term
in~\Ref{nmw2}, describing
the rate of merging, follows from the observation that the smallest
interval between~$n$ points scattered uniformly on a circle of 
circumference~$c$ has an approximately exponential distribution with
mean
$c/n(n-1)$, and that unreached intervals shrink at rate~$2$.  These
equations have an explicit solution $\hat r$ and~$\hat B$, with (for
$\rho>0$)
\bea\label{nmw3}
1-\hat r(Lw)/L = {(a+1)^2 - (a-1)^2 e^{2aL\rho w} \over 2\{
  (a+1) + (a-1)e^{2aL\rho w}\}} =: \hat p(w),
\ena
where $a=\sqrt{1+4/L\rho}$; this is used for
$$
0 \le w \le {1\over aL\rho} \log\{(a+1)/(a-1)\} =: w^*,
$$
the range in which $\hat p(w) \ge 0$.  Then, if~$D$ is the random
variable denoting the distance from~$P$ to a randomly chosen point,
the NMW heuristic takes
$$
\pnmw[D>t] = 1-\hat r(t)/L
$$
as an approximation to the true value $\PP[D>t] = 1- \EE r(t)/L$, where
$r(t)$ denotes the random quantity defined in the growth and merge
model.  The formula for $\EE\ell$ resulting from this heuristic is then
$\frac{L}{k}\int_0^{w^*}\hat p(w)\,dw$ with $\rho=2k\phi$, the factor
$1/k$ arising from the definition of graph distance in the
$k$--neighbour
model, as observed above.

Note that the NMW heuristic always gives $\pnmw[D>Lw^*]=0$.  However,
since the probability of having {\it no\/} shortcuts is $e^{-L\rho/2}$,
it is clear that in fact
$$
\PP[D > Lw^*] > e^{-L\rho/2} (1 - 2w^*) 
     > e^{-L\rho/2} {2 \over 3(1+4/L\rho)},
$$
so that their heuristic cannot give accurate results unless $L\rho$ is
either very small or very big. If~$L\rho$ is very small, then
$\pnmw[D > t] = 1-2t/L + O(L\rho)$, and the same is true for $\PP[D>t]$,
reflecting that there are no shortcuts, except for a probability
of order $O(L\rho)$.  The interesting case is that in which there are
many shortcuts, when $L\rho$ is large, and this we investigate
rigorously.

The early development of~$R$ is close to that of a birth and
growth process~$S$, defined as follows.  We let $(M(t),\,t\ge0)$ be
a Yule (linear Markov pure birth) process with {\it per capita\/}
birth rate~$2\r$, having $M(0)=1$.  To the
$j$'th individual born in the process, $j\ge1$, we associate a
centre~$\z_{j+1}$, where $(\z_j,\,j\ge2)$ are independent and
uniformly distributed on~$C$; we assign $\z_1 := P$ to the initial
individual.  Then, for any $0 \le t \le L/2$, we define~$S(t)$ to
be the set of~$M(t)$ possibly overlapping intervals
$$
S(t) := \{[\z_j-(t-\s_j),\z_j + (t-\s_j)],\ 1\le j\le M(t)\},
$$
where $\s_{j+1}$ denotes the birth time of the~$j$'th individual born,
$j\ge1$, and $\s_1 := 0$.  In fact, such a process~$S$ can be
constructed on the same probability space as~$R$, with differences
arising only when intervals intersect, in the following way.  First,
{\it every\/} potential chord is accepted in~$S$, so that no
thinning takes place, and the chords that were not accepted for~$R$
initiate independent birth and growth processes having the same
distribution as~$S$, starting from their label~2 end points.
Additionally, whenever two intervals intersect,
they continue to grow, overlapping one another; in~$R$, the pair
of end points that meet at the intersection contribute no further
to growth, and the number of intervals in~$R$ decreases by~$1$,
whereas, in~$S$, each end point of the pair continues to generate
further chords according to independent Poisson processes of
rate~$\r$, each of these then initiating further independent
birth and growth processes.

The process~$S$ thus constructed agrees closely with~$R$ until
appreciable numbers of intersections occur.  Its advantage over~$R$
is the inbuilt branching structure, which makes it much easier
to analyze.  In particular, $\EE M(t) = e^{2 \rho t}$, and 
$s(t) := |S(t)| = \int_0^t
2 M(u) du$, so that $\EE s(t) = \rho^{-1} (e^{2 \rho t} -1)$.
Furthermore, $e^{-2 \rho t} M(t) \rightarrow
W$ a.s., where $W \sim NE(1)$, the negative exponential distribution
with mean~$1$, and hence $e^{-2 \rho t} s(t) \rightarrow
\rho^{-1}W$ a.s. and
$\frac{s(t)}{M(t)} \rightarrow \rho^{-1}$ a.s.; note also that~$R(t)$
is contained in the union of the intervals of~$S(t)$, and that
$M(t) \ge B(t)$ a.s.  We make ample use of these facts in the
coming argument.

In order to discuss the distance between two points $P$ and~$P'$,
we modify this construction a little.  We choose two independent
starting points $P$ and~$P'$ uniformly on~$C$, and run two such
constructions $(R,S)$ and~$(R',S')$ simultaneously, based on the
{\it same\/} set of potential chords.  A potential chord $\{q,q'\}$
such that $\partial R(t-) \cap Q = \{q\}$ is only accepted for~$R$
if $q \notin R(t-)\cup R'(t-)$, and it initiates an independent
birth and growth process in~$S$ if not accepted for~$R$; the 
corresponding rule holds if
$\partial R'(t-)\cap Q = \{q\}$.  If two intervals in $R\cup R'$
merge, the pair of end points which meet contribute nothing further 
to $R$ or~$R'$, but each continues to contribute independently to
$S$ or~$S'$, as appropriate.  This construction is actually the
same as the previous, but starting with $M(0) = 2$; further, a
record is kept of which of the two initial individuals was the
ancestor of each subsequent interval.  If, by time~$t$, no pair
of intervals, one in~$R$ and the other in~$R'$, have merged, then
$d(P,P') > 2t$; otherwise, $d(P,P') \le 2t$.

Our strategy is to approximate the event of an $R$--$R'$ merging
of intervals having occurred by looking for $S(t)$--$S'(t)$
intersections.  Every pair of $R$--$R'$ intervals merged up to
time~$t$ is contained in a pair of $S(t)$--$S'(t)$ overlapping
intervals.  However, there may be other $S(t)$--$S'(t)$
overlapping pairs, either (Type~I) because one or other of
the pair arose as progeny in a birth and growth process which
was not part of $R$ or~$R'$ --- following the non--acceptance
of a chord in~$R$ or~$R'$, or the merging of two intervals ---
or (Type~II) because one of the pair is itself an interval
$[q' - (t-\s),q'+(t-\s)]$ coming from a chord $\{q,q'\}$ which
was {\it not\/} accepted at time~$\s$, and the other is an
interval which contained~$q'$ at time~$\s$.  Type~II pairs
can be recognized at time~$t$, because one of the intervals is
entirely contained in the other.

So, taking $L\rho$ large,
consider the situation at time $\tau_x = \frac{1}{2 \rho}
\left\{ \frac{1}{2}\log(L\rho) + x \right\}$, with
$x \geq -\frac{1}{2}\log (L\rho)$ to make $\tau_x \ge 0$.
 Letting $R_x=R(\tau_x), { S}_x = {S}(\tau_x),
M_x = M(\tau_x)$, and $s_x = s(\tau_x)$, we have $\EE M_x = (L
\rho)^{1/2} e^x$ and $\EE s_x = \rho^{-1} \left( (L \rho)^{1/2} 
e^x - 1 \right)$. Thus there are about $L \rho e^{2x}$ 
pairs of intervals with one in ${S}_x$ and the
other in ${S}'_x$, and each is of typical length $\rho^{-1}$, so that
the expected number of intersecting
pairs of intervals  is about $\frac{2}{L \rho} L\rho e^{2x} = 2 e^{2x}$, 
which is small when~$x$ is large and negative, and becomes  
large as~$x$ increases to become large and positive. 
\ignore{in the chosen range of $x$, grows
from $2(L\rho)^{-1/2}$ to $2(L\rho)^{1/2}$.} 
 We show that the
number of Type~I pairs is of rather smaller order, so that their
influence is unimportant.  However, the expected number of Type~II
pairs is
$$
2\int_0^{\t_x} 2\r e^{2\r u}\,L^{-1}\ex s(u)\,du \sim e^{2x},
$$
or about half the total number of intersecting $S(t)$--$S'(t)$
pairs, and these have to be taken into account. Labelling the 
intervals in $S_x$ as $I_1, \ldots, I_{M_x}$ and the intervals in
$S_x'$ as $J_1, \ldots, J_{N_x} $, where the indices are assigned in
chronological order of birth, we set
\bea  \label{xij}
X_{ij} = {\bf 1}\{ I_i \cap J_j \neq \emptyset\}\,
 {\bf 1}\{ I_i \not\subset J_j \}\,
 {\bf 1}\{ J_j \not\subset I_i\},  
\ena
and write
\bea \label{what}
\whv_x = \sum_{i=1}^{M_x} \sum_{j=1}^{N_x} X_{ij}.
\ena
We show that the event
$\whv_x=0$ is with high probability the same as the event $V_x=0$, 
where
$V_x$ is the number of $R_x$--$R'_x$ merged pairs of intervals. 
Finally, if there are no $R_x$--$R'_x$ merged pairs, the ``small worlds''
distance between $P$ and $P'$ is more than $2 \tau_x$. 
For later use, let $p_I(l)$ denote the unique index $j<l$ of the
interval $I_j$ to which $I_l$ is linked by a chord --- the `parent' 
of~$I_l$ --- and define~$p_J$ analogously.

\section{The continuous circle model: proofs}
 \setcounter{equation}{0}

The first step in the argument outlined above is to establish a Poisson
approximation theorem for the number of pairs of $S_x$--$S'_x$
overlapping intervals, excluding Type~I and Type~II pairs.  
The approximation is based on the following general result.

\begin{theorem}\label{Poisson}
Let $X_1,\ldots,X_m$ and $Y_1,\ldots,Y_n$ be independent random
elements,
and let $\fij := \fij(X_i,Y_j)$ be 
indicator random variables with $\pij := \EE \fij$ 
such that
$$
\max\{\max_{k\neq j}\EE(\fij\phi_{ik}),\max_{l\neq i}\EE(\fij\phi_{lj})\} 
\le p\pij,
$$
for all pairs $i,j$, where $p \ge \max_{i,j}\pij$. Then, if 
$\Phi := \siim \sjn \fij$, 
$$
\dtv(\law(\Phi),\Po(\EE \Phi)) \le (2(m+n)-3)p.
$$
\end{theorem}

\proof  Using the local version of the Stein--Chen method (\BHJ,
Theorem~1.A),
we have
$$
\dtv(\law(\Phi),\Po(\EE \Phi)) 
  \le (\EE\Phi)^{-1}\siim\sjn(\pij^2 + \pij\EE \Zij + \EE(\fij\Zij)),
$$
where
$$
\Zij := \sum_{{l=1 \atop l\neq i}}^m \phi_{lj}
    + \sum_{{l=1 \atop l\neq j}}^n \phi_{il}.
$$
By assumption, $\EE(\fij\Zij) \le (m+n-2)p\pij$, and the result
follows.
\ep

This theorem, in the context of intervals scattered on the circle, has
the
following direct consequence.

\begin{corollary}\label{cor1}
Let $m$ intervals $I_1,\ldots,I_m$ with lengths $s_1,\ldots,s_m$ and
$n$ intervals $J_1,\ldots,J_n$ with lengths $u_1,\ldots,u_n$ be
positioned
uniformly and independently on~$C$.  Set $\whv:= \siim\sjn \Xij$, where
$$
\Xij := I[I_i\cap J_j \neq \emptyset]\, I[I_i \not\subset J_j]\,
  I[J_j \not\subset I_i]. 
$$
Then
$$
\dtv(\law(\whv),\Po(\l_{(m,n,\bs,\bu)})) \le 4(m+n)l_{\bs\bu}/L,
$$
where 
\beas
&&\l_{(m,n,\bs,\bu)} := 2L^{-1}\siim\sjn\min\{s_i,u_j\}, \quad
\bs:= (s_1,\ldots,s_m),\\
&&  \bu := (u_1,\ldots,u_n) 
\quad \mbox{{and}}\quad l_{\bs\bu} := \max\{\max_i s_i,\max_j u_j\}.
\enas
\end{corollary}

\proof  We apply \Tr{Poisson} with $X_i$ and~$Y_j$ the centres of the
intervals $I_i$ and~$J_j$, and with $\Xij$ for~$\fij$; the~$\Xij$ are
pairwise
independent, and satisfy $\EE\Xij \le 2l_{su}L^{-1}$.  It remains only
to note that 
$$
\EE \whv = 2L^{-1}\siim\sjn \min\{s_i,u_j\} = 
\l_{(m,n,\bs,\bu)}.\eqno\Box
$$
 
\bigskip
The corollary translates immediately into a useful statement about
$\whv_x$.
We define $\bs_x$ and~$\bu_x$ to be the sets of lengths of the
intervals of $S_x$ and~$S'_x$, and we always take $x\ge -\half\log(L\rho)$,
so that $\tau_x \ge 0$.

\begin{corollary}\label{cor2}
For the processes $S$ and~$S'$ of the previous section, we have
\beas
&&|\PP[\whv_x = 0\giv M_x=m,N_x=n,\bs_x=\bs,\bu_x=\bu] -
   \exp\{-\l_{(m,n,\bs,\bu)} \}|\\
&&\qquad \le 8(m+n)\tau_x L^{-1}.
\enas
\end{corollary}

\proof  It suffices to note that all the intervals of $S_x$ and~$S'_x$
are of length at most $2\tau_x$, and hence that the chance of a
given pair intersecting is at most $4L^{-1}\t_x$.
\ep

\remark  If $P$ and~$P'$ are not chosen at random, but are fixed points 
of~$C$,
the result of Corollary~\ref{cor2} remains essentially unchanged,
provided that they are more than an arc distance of~$2\tau_x$
apart.  The only difference is that then $X_{11}=0$ a.s., and that
$\l_{(m,n,\bs,\bu)}$ is replaced by $\l_{(m,n,\bs,\bu)} - 4\tau_x/L$.  
If $P$ and~$P'$ are less
than~$2\tau_x$ apart, then $\PP[\whv_x=0] = 0$.

\medskip

The next step is to show that $\PP[\whv_x=0]$ is close to $\PP[V_x=0]$.
We do this by directly comparing the random variables $\whv_x$ and~$V_x$ 
in the joint construction.

\begin{lemma}\label{cor3}
With notation as above, we have
\beas
\PP[\whv_x \neq V_x] &\le& 4\tau_x L^{-2}\EE\{
      (M_xu_x + N_xs_x)(M_x+N_x-1)(1+\log M_x + \log N_x)\}.
\enas
\end{lemma} 

\proof  Let the intervals of $S_x$ and~$S'_x$ and their lengths
be denoted as for Corollary~\ref{cor1}, with $m$ and~$n$
replaced by $M_x$ and~$N_x$, and write 
$s_x := \siimx s_i$ and $u_x := \sjnx u_j$ as before; 
set $S_x^i := \bigcup_{l\neq i} I_l$, $S_x'{}^j := 
\bigcup_{l\neq j} J_l$.	 Define
\beas
H_{i1} &:=& \{I_i\cap S_x^i \neq\emptyset\};\qquad
H_{i2} := \bigcup_{1\le l < i} \Bl\{I_l\cap S_x^l \neq \emptyset\}
  \cap\{A^l(i) = l\}\Br\\
H_{i3} &:=& \bigcup_{1\le l < i} \Bl\{I_l\cap S_x' \neq \emptyset\}
  \cap\{A^l(i) = l\}\Br,
\enas
where
\ignore{, considering $\{1,2,\ldots,l\}$ as the set of ancestors,}
$A^l(i)\in\{1,2,\ldots,l\}$ denotes 
\ignore{which of them was the}
the largest of these indices to be an ancestor
of~$i$, and set $H_i := \bigcup_{v=1}^3 H_{iv}$; define $H'_{iv}$,
$1\le v\le 3$ and~$H'_i$ analogously. Note that, for $l<i$, the event
$\{A^l(i) = l\}$ is the event that~$l$ is an ancestor of~$i$.
Then, with $X_{ij}$ defined
as in~\Ref{xij},	
\bea \label{2.18}
{\whv_x} \geq V_x \geq \sum_{i=1}^{M_x} \sum_{j=1}^{N_x}
X_{ij}I[H_i^c]\,I[H_j'{}^c],
\ena 
so that
$$  
0 \le {\whv_x} - V_x \leq \sum_{i=1}^{M_x} \sum_{j=1}^{N_x}
X_{ij}(1-I[H_i^c]\,I[H_j'{}^c]),
$$ 
and hence 
\bea 
&&\PP [\whv_x \neq V_x ] \leq \EE(\whv_x-V_x)
%  \phantom{HHHHHHHHHHHHHHHHHHHHH}
	 \label{2.19}\\
&&\qquad\le   \EE \left\{ \sum_{i=1}^{M_x} \sum_{j=1}^{N_x}  X_{ij}
     \sum_{v=1}^3(I[H_{iv}=1] + I[H_{jv}'=1]) \right\}. 
		 \non
\ena
Now, conditional on $M_x$, $N_x$, $\bs_x$ and~$\bu_x$, the 
indicator~$X_{ij}$
is (pairwise) independent of each of the events $H_{iv}$ and~$H'_{jv}$,
$1\le v\le 3$, because $H_{i1}$, $H_{i2}$ and~$H'_{j3}$ are each
independent of~$\z'_j$, the centre of~$J_j$, and $H'_{j1}$, $H'_{j2}$
and~$H_{i3}$ are each independent of~$\z_i$.  Moreover, the event
$\{A^l(i) = l\}$ is independent of $M$, $N$ and all $\z_i$'s 
and~$\z'_j$'s, and has probability~$1/l$, since the~$l$ Yule
processes, one generated from each interval~$I_v$, $1\le v\le l$,
which combine to make up~$S$ from time~$\s_l$ onwards, are
independent and identically distributed.  Hence, observing also
that no interval at time~$\t_x$ can have length greater than~$2\t_x$,
it follows that
\beas
&&\EE\{X_{ij}I[H_{i1}] \giv M_x, N_x,\bs_x,\bu_x\}
  \le L^{-1}(s_i+u_j)(M_x-1)(4\t_x/L);\\
&&\EE\{X_{ij}I[H_{i2}] \giv M_x, N_x,\bs_x,\bu_x\}
  \le \sum_{l=1}^{i-1}l^{-1}\,L^{-1}(s_i+u_j)4\t_x(M_x-1)/L\\
&&\qquad \le 4L^{-2}\t_x(M_x-1) \log M_x (s_i+u_j),
\enas
and
\beas
&&\EE\{X_{ij}I[H_{i3}] \giv M_x, N_x,\bs_x,\bu_x\}
  \le \sum_{l=1}^{i-1}l^{-1}\,L^{-1}(s_i+u_j)4\t_xN_x/L\\
&&\qquad \le 4L^{-2}\t_xN_x \log M_x (s_i+u_j).
\enas
Adding over $i$ and~$j$ thus gives	
\beas
\lefteqn{\siimx\sjnx \EE\{X_{ij}I[H_i]\}}\\
&\le& 4L^{-2}\t_x \EE\Blb [(M_x-1)(1+\log M_x) + N_x\log M_x]
  (M_xu_x + N_xs_x)\Brb,
\enas
and combining this with	the contribution from $\EE\{X_{ij}I[H_i']\}$
completes the proof.
\ep

To apply Corollary~\ref{cor2} and Lemma~\ref{cor3}, it remains to establish
more detailed information about the distributions of $M_x$ and~$s_x$.
In particular, we need to bound the first and second moments of~$M_x$,
and to approximate the quantity $\EE\exp\{-\l_{(M_x,N_x,\bs_x,\bu_x)}\}$,
where
\bea
\l_{(M_x,N_x,\bs_x,\bu_x)} &=& 2L^{-1}\siimx\sjnx \min(s_i,u_j)\non\\
&=& 4L^{-1}\int_0^{\t_x}\int_0^{\t_x} \min(\t_x-v,\t_x-w) 
         dM(v)dN(w)\non\\
&=& 4L^{-1}\int_0^{\t_x} M(v)N(v)\,dv. \label{new-2}
\ena	
As from now, we assume that $L\rho \ge 16$.
We begin with the following lemma.

\begin{lemma} \label{Lemma1}
For any~$x \ge -\half\log(L\rho)$, we have 
$$
\PP[M_x \ge r] = (1-q_x)^{r-1},\quad r\ge1,
$$
where $q_x = e^{-2\rho\tau_x}$. Hence, in particular,  
\bea \label{2.01}
\EE M_x = e^{2 \rho \tau_x } = e^x \sqrt{L\rho}, \qquad \tfrac{1}{2} \EE
\{M_x (M_x+1)\} = e^{2x}L \rho
\ena
and
\[
\tfrac{1}{6} \EE \{M_x (M_x+1)(M_x+2)\} = \Bl e^{x}\sqrt{L \rho}\Br^3,
\]
and, if $x \le \quarter\log(L\r)$,
\bea 
\EE\{\log M_x\} &\le& \tfrac54\log(L\r);\label{new-2.2}\\
\EE\{M_x\log M_x\}  &\le& \tfrac74\log(L\r)\,e^x\sqrt{L\r},\label{new-3}
\ena
and
\eq\label{new-3.2}
\EE\{M_x(M_x+1)\log M_x\}  \le \tfrac92\log(L\r)\,e^{2x}L\r;
\en
furthermore, 
\bea \label{2.02}
\EE s_xM_x \le 2\r^{-1}e^{2x}L\r. 
\ena
\end{lemma}

\proof Let $z_t=z_t(w):=\EE \left(w^{M(t)}\right)$. 
Then, by splitting at the first jump, which is 
exponentially distributed with mean $1/2\rho$, we have
\bea \label{2.1}
z_t = w e^{-2 \rho t} + \int_0^t 2 \rho e^{-2 \rho u}
            z^2_{t-u} \, du.
\ena
Multiplying by $e^{2\rho t}$ and differentiating, it follows that 
\bea \label{2.2}
\frac{dz}{dt} = - 2 z \rho(1 - z); \hh z_0=w. 
\ena
Solving the differential equation now gives 
\bea \label{2.4}
\EE \left( w^{M(t)} \right) = w e^{-2 \rho t}\left(1 - w(1-e^{-2 \rho t
}) \right)^{-1}, 
\ena 
so that
\bea \label{2.5}
\PP[M(t)=m] = e^{-2 \rho t}\left(1-e^{-2\rho t} \right)^{m-1}, \hh m
\geq 1.
\ena
The moments in~\Ref{2.01} are immediate, 
and~\Ref{new-2.2}--\Ref{new-3.2} follow
because $\log m \le 2\r t + me^{-2\r t}$, so that, for instance,
\beas
\EE\{M_x\log M_x\} &\le& 2\r\t_x \EE M_x + \EE M_x^2e^{-2\r\t_x}\\
 &\le&  (\tfrac34 \log(L\r) + 2)e^{2\r\t_x},
\enas
and $\log(L\r) \ge 2$ in $L\r \ge 16$. Finally,
\beas
\EE\{s(t)M(t)\} &=& 2\int_0^t \EE\{M(u)M(t)\}\,du\\
  &=& 2\int_0^t e^{2\r(t-u)}\EE\{M^2(u)\}\,du\\
	&=& 2\r^{-1}e^{4\r t}\{1 - e^{-2\r t}(1+\r t)\},
\enas
and the lemma is proved.	
\hfill $\Box$

\medskip
We shall also need some information about the conditional distribution
of~$M(s)$ given $\ffti := \s\{M(u),\,u\ge t\}$, for $s < t$.
This is summarized in the next lemma.

\begin{lemma} \label{conditional}
For any $0\le s\le t$,
$$
\law(M(s) \giv \ffti\cap\{M(t)=m\}) = 1 
   + \Bi\Bl m-1,\frac{e^{-2\r(t-s)}(1-e^{-2\r s})}{1-e^{-2\r t}}\Br.
$$
In particular, setting $W(u) := e^{-2\r u}M(u)$, it follows that
\bea\label{W-condmean}
\EE(W(s)-W(t) \giv \ffti) = \Bl \frac{e^{-2\r s} - e^{-2\r t}}
  {1-e^{-2\r t}} \Br (1-W(t))
\ena
and
\beas
\lefteqn{\EE\{(W(s)-W(t))^2 \giv \ffti\}}\\
 &=& \Bl \frac{e^{-2\r s} - e^{-2\r t}}
  {(1-e^{-2\r t})^2}\Br \{(e^{-2\r s} - e^{-2\r t})(1-W(t))^2
	  + (W(t) - e^{-2\r t})(1 - e^{-2\r s})\}.
\enas	
Furthermore, 
\eq\label{new-5.1}
\EE\{s_x\log M_x\} \le \tfrac74 \r^{-1}\log(L\r)(1+e^x)\sqrt{L\r}
\en
and	
\eq\label{new-5.2}
\EE\{s_xM_x\log M_x\} \le \tfrac92 \r^{-1}\log(L\r)e^x(1+e^x)L\r.
\en
\end{lemma}

\proof  From the branching property of~$M$, and with~$z_u$ as in
the previous lemma, it follows that
\beas
\EE\{v^{M(s)}w^{M(t)}\} &=& \EE\{v^{M(s)}(z_{t-s}(w))^{M(s)}\}
  = z_s(vz_{t-s}(w)) \\
&=& \frac{vwe^{-2\r t}}
   {1 - w(1-e^{-2\r(t-s)}) - vw	e^{-2\r(t-s)}(1-e^{-2\r s})}.
\enas
Hence, from the coefficient of~$w^m$, and since~$M$ is a Markov process, 
the probability generating function of 
$\law(M(s) \giv \ffti\cap\{M(t)=m\})$ is
$$
v\Blb \frac{(1 - e^{-2\r(t-s)}) + ve^{-2\r(t-s)}(1-e^{-2\r s})}
  {1-e^{-2\r t}} \Brb^{m-1},
$$
proving the first statement of the lemma. The $W$--moments
are now immediate. For the last part, note that
\beas
\EE(s_x \giv M_x) &=& 2\int_0^{\t_x} \ex(M(u)\giv M_x)\,du\\
&\le& 2\int_0^{\t_x}\Blb e^{2\r(u-\t_x)}M_x(1-e^{-2\r\t_x})^{-1}
  + 1\Brb\,du\\
&\le& \r^{-1}M_x + 2\t_x,
\enas
again using the first part of the lemma, and the remaining
conclusions follow from Lemma~\ref{Lemma1}	
\hfill $\Box$

\medskip	
The estimates of Lemmas \ref{Lemma1} and~\ref{conditional} 
can already be applied to Corollary~\ref{cor2} and Lemma~\ref{cor3}. 

\begin{corollary}\label{cor4}
For any $x\ge -\half\log(L\rho)$, we have
\beas
&&|\PP[V_x = 0] - \EE\exp\{-\l_{(M_x,N_x,\bs_x,\bu_x)}\}| \\
&&\qquad\le 6e^x(L\rho)^{-1/2}\log(L\rho) +
           60e^{2x}(1+e^x)(L\rho)^{-1/2}\log^2(L\rho).
\enas
\end{corollary}

We now estimate the quantity $\EE\exp\{-\l_{(M_x,N_x,\bs_x,\bu_x)}\}$.
Since $e^{-2\rho t}M(t)$ is a martingale, and converges a.s.~to a
limit~$W$,
and since $e^{2\rho\tau_x} = e^x\sqrt{L\rho}$, it is clear 
from~\Ref{new-2} that
$$
\exp\{-\l_{(M_x,N_x,\bs_x,\bu_x)}\} \sim 
   \exp\Blb-4L^{-1}WW'\int_0^{\t_x} e^{4\r v}\,dv \Brb
   \sim \exp\{-e^{2x}WW'\},
$$
where~$W'$ is an independent copy of~$W$, 
so that $\PP[V_x=0]$ can be approximated in terms of the distribution of
the limiting random variable~$W$ associated with the Yule process~$M$.
Here, we make the precise calculations.

\begin{lemma} \label{Lemma2}
We have 
\beas
\lefteqn{\Blm\EE \left\{ e^{-\l_{(M_x,N_x,\bs_x,\bu_x)}}\right\} 
    - \int_0^\infty \frac{e^{-y}}{1 + c_xy}dy \Brm}\\
&=& O\left\{e^x(1+e^{2x})(L\rho)^{-\frac{1}{2}} \right\},
\enas 
uniformly in $-\half\log(L\rho) \leq x  \leq \frac{1}{4} \log(L \rho)$, where
$c_x=e^{2x}$. \end{lemma}

\bigskip
\remark 
Recall that $W = \lim_{t \rightarrow \infty} e^{-2 \rho t}M(t)$ and
$W' = \lim_{t \rightarrow \infty} e^{-2 \rho t}N(t)$. Note that, as $W$
and $W'$ are independent and exponentially distributed with mean~$1$, 
we have 
\bea\label{Wform} 
\int_0^\infty \frac{e^{-y}}{1 + c_xy}dy = \EE \left( e^{-W W' e^{2x}}
\right).
\ena

\medskip

\proof
We begin by observing that, for $a,b>0$,
$$
|e^{-a} - e^{-b} - (b-a)e^{-b}| \le \half(b-a)^2.
$$
Hence it follows that
\bea
&&\Blm \EE\Bl \exp\Blb -\frac4L \int_0^t M(u)N(u)\,du \Brb
  -  \exp\Blb -\frac4L \int_0^t We^{2\r u}N(u)\,du \Brb
	    \right.\right.\non\\
&&\qquad\quad \left.\left.\mbox{} 
    + \frac4L \int_0^t(M(u)-We^{2\r u})N(u)\,du\,
  	 \exp\Blb -\frac4L \int_0^t We^{2\r v}N(v)\,dv \Brb\Br\Brm\non\\
&&\quad\le \frac12\EE\Blb \frac4L \int_0^t(M(u)-We^{2\r u})N(u)\,du \Brb^2,
    \label{new-4}
\ena
and also that
\bea
&&\Blm \EE\Bl \exp\Blb -\frac4L \int_0^t We^{2\r u}N(u)\,du \Brb
  -  \exp\Blb -\frac4L \int_0^t WW'e^{4\r u}\,du \Brb
	    \right.\right.\non\\
&&\qquad\quad \left.\left.\mbox{} 
    + \frac4L \int_0^t We^{2\r u}(N(u)-W'e^{2\r u})\,du\,
  	 \exp\Blb -\frac4L \int_0^t WW'e^{4\r v}\,dv \Brb\Br\Brm\non\\
&&\quad\le \frac12\EE\Blb 
   \frac4L \int_0^tWe^{2\r u}(N(u)-W'e^{2\r u})\,du \Brb^2.
    \label{new-5}
\ena

Now, examining the second line of~\Ref{new-4}, we first condition
on~$W$, which is equivalent to conditioning on~$W(t)$ for very
large~$t$, and apply~\Ref{W-condmean} from
Lemma~\ref{conditional} to give
\beas
\lefteqn{\frac4L \EE \Bl \int_0^t(M(u)-We^{2\r u})N(u)\,du\,
  	 \exp\Blb -\frac4L \int_0^t We^{2\r v}N(v)\,dv \Brb \Br}\\
&=& \frac4L \EE \Bl \int_0^t(1-W)N(u)\,du\,
  	 \exp\Blb -\frac4L \int_0^t We^{2\r v}N(v)\,dv \Brb \Br;
\enas
hence it follows that
\beas
\lefteqn{\frac4L \Blm \EE \Bl \int_0^{\t_x}(M(u)-We^{2\r u})N(u)\,du\,
  	 \exp\Blb -\frac4L \int_0^{\t_x} We^{2\r v}N(v)\,dv \Brb \Br \Brm}\\		 
&\le& \EE\Blb|1-W|\,\frac4L \int_0^{\t_x} N(u)\,du\Brb 
   \le \frac2{L\r} e^{2\r \t_x} = 2e^x(L\r)^{-1/2}.
\enas
For the last line of~\Ref{new-4}, we have
$$
\EE\{(M(u) - We^{2\r u})(M(v) - We^{2\r v})\} 
    = e^{2\r(u+v)}\EE\{(W(u)-W)(W(v)-W)\}.
$$
Writing 
$$
c(s,t) := \frac{e^{-2\r s} - e^{-2\r t}}{1-e^{-2\r t}}, 
$$
and again using Lemma~\ref{conditional},
it thus follows for $0 < u < v$ that
\beas
\lefteqn{\EE\{(W(u)-W)(W(v)-W)\}}\\
&=& c(u,v)\EE\{(1-W(v))(W(v)-W)\} + \EE\{(W(v)-W)^2\}   \\
&=& (1-c(u,v))\EE\{(W(v)-W)^2\} + c(u,v)\EE\{(1-W)(W(v)-W)\} \\
&=& e^{-2\r v}(1-c(u,v))\Blb \EE W(1-e^{-2\r v}) 
   + e^{-2\r v}\EE\{(1-W)^2\}\Brb \\
&&\qquad\mbox{}	 + c(u,v)e^{-2\r v}\EE\{(1-W)^2\},
\enas
so that $0 \le 	\EE\{(W(u)-W)(W(v)-W)\} \le Ke^{-2\r v}$ for some
$K > 0$. Hence
\beas
&&\EE\Blb \frac4L \int_0^{\t_x}(M(u)-We^{2\r u})N(u)\,du \Brb^2\\
&&\qquad = \frac{16}{L^2} \int_0^{\t_x} \int_0^{\t_x} e^{2\r (u+v)}
  \EE\{(W(u)-W)(W(v)-W)\} \EE\{N(u)N(v)\}\,dudv\\
&&\qquad \le \frac{32}{L^2} \int_0^{\t_x} \int_0^{v} Ke^{2\r u}
  \EE\{N(u)N(v)\}\,dudv,
\enas
and, for $0 < u < v$,
$$
\EE\{N(u)N(v)\} = e^{2\r(v-u)}\EE N^2(u) \le 2e^{2\r (u+v)},
$$
from~\Ref{2.01}; this gives	
$$
\EE\Blb \frac4L \int_0^{\t_x}(M(u)-We^{2\r u})N(u)\,du \Brb^2
  \le (8K/3)e^{3x}(L\r)^{-1/2}
$$			 			 		 	 
for the last line in~\Ref{new-4}.  A similar argument for the
components of~\Ref{new-5} completes the proof.				 
\ep

\begin{theorem}\label{theorem5}
If $P$ and~$P'$ are randomly chosen on~$C$, or if $P$ and~$P'$ are 
fixed points of~$C$ at arc distance more than $\frac3{4\rho}\log(L\rho)$ 
from one another, then 
\beas
\lefteqn{ \left\vert \PP \left[ D>\frac{1}{\rho}
\left(\frac{1}{2} \log(L \rho) + x \right) \right] 
- \int_0^\infty \frac{e^{-y}dy}{1+c_xy}\right\vert  }\\
&=&O \left( e^{x}(1+e^{2x}) (L\rho)^{-\frac{1}{2}}\log^2(L\rho) \right),
\enas
uniformly in $-\half\log(L\rho) \le x \le \frac14\log(L\rho)$, where, 
as before, $D$ denotes
the shortest distance between $P$ and~$P'$ on the shortcut graph.
\end{theorem}

\proof  Since 
$$
\{V_x=0\} = \left\{ D >  \frac{1}{\rho}\left(\frac{1}{2} 
       \log(L \rho) + x \right) \right\},
$$
the theorem follows from Corollary~\ref{cor4} and 
Lemma~\ref{Lemma2}. 
\ep

\begin{corollary}\label{corollary6}
If $T$ denotes a random variable with distribution given by
$$
\PP\left[T > x \right]
= \int_0^\infty \frac{e^{-y}dy}{1+e^{2x}y} 
$$
and $D^* =  \frac{1}{\rho}\left(\frac{1}{2} \log(L \rho) + T \right)$,
then
\beas
\sup_x|\PP[D\le x] - \PP[D^*\le x]| 
   = O\left( (L \rho)^{-\frac{1}{5}} \log^2(L \rho) \right). 
\enas
\end{corollary}

\proof 
%As $c \rightarrow 0$, $ \int_0^\infty \frac{e^{-y}dy}{1+cy} \sim 1-c$; 
As $c \rightarrow \infty$, we have 
$$
\int_0^\infty \frac{e^{-y}dy}{1+cy} 
       \leq \left(1 - e^{-\frac{1}{c}}  \right) + \frac{1}{c}
 \log c + \frac{1}{c} \sim \frac{1}{c} \log c.
$$
So use the bound from Theorem \ref{theorem5} for 
%$-\frac{1}{6} \log(L \rho) \leq 
$x \leq \frac{1}{10} \log (L \rho)$, 
and the tail estimates of ${\cal L}(D^*)$ outside this range. 
\ep

\begin{figure}[h]
\centerline{\psfig{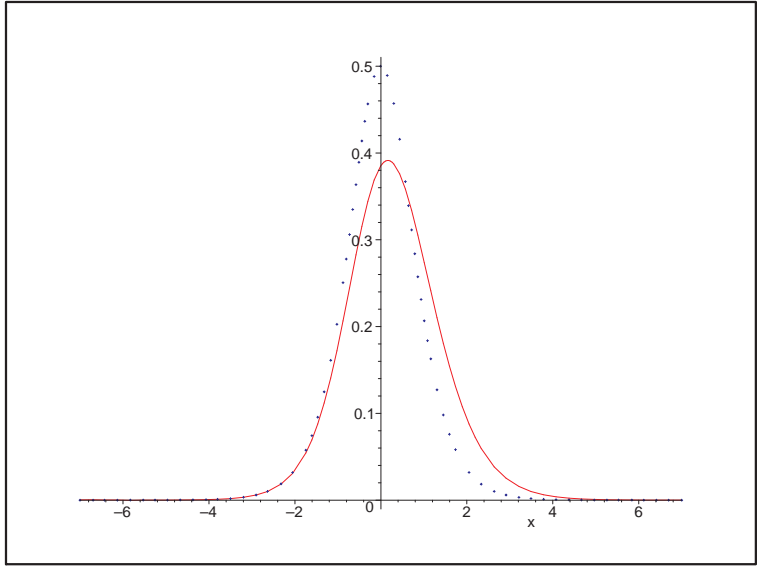}}
\caption{The asymptotic distribution function of
$\rho D - \frac12\log(L\rho)$ (solid line) and that predicted by the
NMW heuristic (dotted line)}
\label{fig1}
\end{figure}

Note that
\ignore{, in $|x| \le \frac14\log(L\rho)$, }
the asymptotics of the NMW heuristic give
$$
\pnmw\left[D > \frac1\rho \left(\frac{1}{2} \log(L \rho) + x
\right)\right]
 = {1\over 1+e^{2x}} ( 1 + O(\{L\rho\}^{-1/2})),
$$
agreeing with \Tr{theorem5} only at the $\log(L\rho)$ order.  Under the
NMW heuristic, the asymptotic distribution of $\rho D-\frac12\log(L\rho)$
is a logistic distribution with mean zero, whereas the true asymptotic
distribution given in Corollary~\ref{corollary6} has mean 
$\EE T = \half \gamma\approx 0.2886$, where~$\gamma$ is Euler's constant,
and has a much wider spread: see Figure~\ref{fig1}.

As in the NMW model, we could instead have taken the number of shortcuts
to be fixed at $\half L\rho$. Using a standard deviation 
argument, such a process could with high probability
be bracketed by two processes with Poisson numbers of shortcuts, but 
with different shortcut rates  
$$ 
\rho_1 = \rho \{ 1 - \kappa L^{-1}\sqrt{L\rho}\log(L\rho)\}
\quad\mbox{and}\quad
       \rho_2 = \rho \{ 1 + \kappa L^{-1}\sqrt{L\rho}\log(L\rho)\}, 
$$
for $\kappa$ big enough. Expanding $\log(L\rho_1)$ and $\log(L\rho_2)$ around $\log(L\rho)$,  we see that the distributional approximation for the 
shortest path length would remain 
the same, for a slightly different region and  with different error bounds.

\section{$r$ dimensions}
 \setcounter{equation}{0}

Our method of proof can be adapted to many other models.  Here,
we consider only the generalization of the previous continuous circle
model to higher dimensions, taking $\Po(L\rho/2)$  shortcuts between
random
pairs of points in a finite, homogeneous space~$C$ in~$r$ dimensions,
such as a sphere or a torus, where $L$ is the area of~$C$.  We construct the
shortcuts by a ``growth and merge'' process as before, 
but now with intervals replaced by local neighbourhoods of the form 
$\z + tK$ after growth time $t$, where~$\z$ is the centre and
$K \ni 0$ is a given convex set in~$r$ dimensions. 
The basic Poisson approximation of \Tr{Poisson} can be applied as
before; we then need to find an appropriate~$\tau_x$, and to
make the necessary computations for the associated pure growth process.
In particular, we shall need to be able to approximate the sums
\bea \label{1}
\sum_{i=1}^{m} \sum_{j=1}^{n} p(s_i, u_j),
\ena
where $p(s_i, u_j)$ is the probability that two independently and
randomly distributed sets, 
one an $s_iK$ and the other a $u_jK$, intersect one another. 
We assume this probability to be of the form 
\bea\label{1+1}
p(s_i,u_j) = L^{-1}\sum_{l=0}^r \alpha_l s_i^l u_j^{r-l} \quad 
\mbox{ for constants\ }
\alpha_l=\alpha_{r-l}.
\ena
In one dimension, as in the previous
section, $\a_0=\a_1=2$; for a torus in two dimensions with~$K$ a unit
square $[-1,1]^2$,
$\a_0 = \a_1/2 = \a_2 = 4$.  For a sphere in two dimensions, it is
almost
the case, neglecting curvature, that $\a_0 = \a_1/2 = \a_2 = \pi$, and
the
error in using this approximation is negligible for large~$L$, to our
order of approximation.  As in 1--dimension, we shall also have to
discount intersections of neighbourhoods where one is entirely
contained in the other.

For the pure growth process with independently and uniformly positioned 
neighbourhoods, define its neighbourhood size process by
the purely atomic measure $(\xi_t,\,t\ge0)$ on $\RR_+$: 
$$
\xi_t(A) = \# \{\mbox{neighbourhoods with
radii having lengths in\ } A \}. 
$$
Then the quantities
$$
M_l(t) := \int_{\RR_+} x^l \xi_t(dx) ,\quad l\ge0,
$$
are basic to our analysis: $M_0(t)$ is just the number of neighbourhoods
in the pure growth process at time~$t$, corresponding to~$M(t)$ in the
circle model, and 
$$
M_l(t) = \sum_{j=1}^{M_0(t)} s_j^l
$$ 
is the sum of the $l$'th powers of the `radii' of the neighbourhoods.
In particular, in view of \Ref{1} and~\Ref{1+1}, the analogue of
$\l_{(m,n,\bs,\bu)}$ of Corollary~\ref{cor1} is
%$$
%L^{-1}\sum_{l=0}^r \a_l \siim s_i^l\sjn u_j^{r-l} - ,
%$$
easily expressible for two pure growth processes $\bM$ and~$\bN$ at
time~$t$ as
\bea\label{1+mean}
L^{-1}\Blb \sum_{l=0}^r \a_l M_l(t)N_{r-l}(t) 
      - v(K)\int_0^t(N_r(u)dM_0(u) + M_r(u)dN_0(u)) \Brb,
\ena
where $v(K)$ is the volume of~$K$. 

The quantities $M_l(t)$, $0\le l\le r$, satisfy the following evolution
equations:
\bea
&&\frac{d}{dt} M_i(t) = i M_{i-1}(t), \quad i\ge 2; \non\\
&&\frac{d}{dt} M_1(t) = M_0(t) \quad \mbox{for a.e. }t; \label{Meqns}\\
&&M_0(t) - \rho rv(K) \int_0^t M_{r-1}(u) du = X_0(t), \non
\ena
where
$X_0$  is a martingale, with $X_0(0)=1$ and with
(centred) Poisson innovations having rate $\rho rv(K) M_{r-1}(t)$ at
time~$t$.  
Properties of their solution are given in the following theorem.

 \begin{theorem} \label{r-D1}
 Let $\bM(t)$ denote the $(r+1)$-vector $(M_0(t), \ldots, M_r(t))^T$.
Then, as $t\to\infty$, 
\beas
\EE \bM(t) \sim r^{-1}e^{\l_0 t} \be\uo;
  \quad \EE M_0^2(t) = O(e^{2\l_0 t}),
\enas
where
\beas
 \lambda_0 = \l_0(\rho) := (r! \rho v(K))^{\frac{1}{r}}; \ 
\ \be\uo = (1, \lambda_0^{-1}, 2 \lambda_0^{-2},
 \ldots, (r-1)! \lambda_0^{-(r-1)}, r!\l_0^{-r})^T.
 \enas
Furthermore, 
 \beas
 \bW(t) &:=& \bM(t) e^{-\lambda_0t} \rightarrow \be\uo W \mbox{ a.s.}, 
 \enas
 where 
 \beas
 W := \frac{1}{r}\left\{1 + \int_0^\infty e^{-\lambda_0 y} dX_0(y)
\right\} 
 \enas
 satisfies $\PP[W > 0] = 1$, and
\beas
(j!)^{-1}\l_0^j\EE|W_j(t)-W\be\uo_j| \le c_r\Bl 1+(\l_0t)^{1/2}1_{\{r=6\}}
  \Br e^{-\l_0(1-\w_*)t},\quad 0\le j\le r,
\enas
for some constant~$c_r$ not depending on~$\rho$, where
$$
\w_* = \max\{0,\cos(2\pi/r)\}.
$$
\end{theorem}

\remark  The fact that $W_0(t) \to W$ a.s.\ implies that
$$
M_0(t)^{-1} \xi_t\{[0,b)\} = 1 - W_0(t-b)e^{-\l_0b}/W_0(t) \to
1-e^{-\l_0b}
$$
for all $b>0$, so that the form of $\be\uo$ is not surprising.

\medskip 

\proof  Solving \Ref{Meqns} for the vector $\bM(t)$, we find that
\bea\label{1+4}
\bM(t) = A e^{At} \int_0^t e^{-Ay} X_0(y) {\bf \e^0} dy + X_0(t) {\bf
\e^0},
\ena
where ${\bf \e^0}$ is the coordinate vector $(1,0,0,\ldots,0)^T$, and 
\beas
A := \left( \begin{array}{ccccccc}
0 &  &  & & 0& c&0\\
1 &0 &  & &  &   & \\
0 & 2 & \cdot& &  &    &\\
\cdot & & & \cdot&&&\cdot\\
\cdot & & & & \cdot &&\cdot\\
\cdot & & & &  &\cdot &\cdot\\
0 & &&& 0 & r & 0
\end{array}
\right)
\enas 
has eigenvalues $\l_r=0$ and $\l_0,\ldots,\l_{r-1}$ satisfying the
equation
 $\lambda^r = (r-1)!c$: here, $c=\rho rv(K)$. Thus $\lambda_0=
\{c(r-1)!\}^{\frac{1}{r}}$ is real and positive, and 
$\lambda_l = \lambda_0 \w_l$, where $\w_l = e^{2 \pi i l / r}$ are the
complex~$r$'th roots of unity.

The eigenvectors $\be\ul$ of~$A$ are also easy to determine.  The $r$'th 
eigenvector $\be^{(r)}$ is the coordinate vector ${\bf \e^r} = 
(0,\ldots,0,1)^T$, and all others have components $e_0,\ldots,e_r$
satisfying the equations
\beas
c e_{r-1}& =&  \lambda e_0; \quad
e_0 = \lambda e_1;\quad
2 e_1 = \lambda e_2;\quad\ldots;\\
(r-1) e_{r-2} &=& \lambda e_{r-1};\quad
r e_{r-1} = \lambda e_r,
\enas 
for $\l = \l_0,\ldots,\l_{r-1}$. These in turn give 
\beas
e_1 & =& \lambda^{-1} e_0;\quad
e_2 = 2 \lambda^{-2} e_0;\quad\ldots;\\
e_{r-1} &=& (r-1)! \lambda^{-(r-1)} e_0 = \lambda c^{-1} e_0;\quad
e_r= r! \lambda^{-r} e_0.
\enas
Thus it follows that, for $ 0\le l\le r-1$,
\bea\label{1+2}
{\bf e}\ul = \left(1, (\lambda_0\w_l)^{-1}, 2 (\lambda_0 \w_l)^{-2},
\ldots, 
   (r-1)! (\lambda_0\w_l)^{-(r-1)}, c^{-1}r\right)^T.
\ena

The eigendecomposition can now be used to determine $\bM(t)$ more
explicitly.
Writing ${\bf \e^0}$ in terms of the $\be\ul$, $0\le l\le r$, we find
that
$$
{\bf \e^0} = \sum_{l=0}^r \mu_l \be\ul\ \ \mbox{with }\  
     \m_0=\cdots=\m_{r-1}=r^{-1}\ \mbox{and } \m_r = -r/c.
$$
Using this to evaluate \Ref{1+4}, we obtain 
\bea\label{1+5}
\bM(t) &=& \int_0^t \{(X_0(y)-1)+1\} \sum_{l=0}^{r-1} \be\ul\mu_l
\lambda_l   
       e^{\lambda_l(t-y)} dy + X_0(t) {\bf \e^0}.
\ena

Turning to the moments of~$\bM$, we immediately have
\beas
\EE \bM(t) &=& \int_0^t  \sum_{l=0}^{r-1}\be\ul \mu_l \lambda_l 
    e^{\lambda_l(t-y)} dy +{\bf \e^0}\\
 &=& \sum_{l=0}^{r-1} \frac{1}{r} \left( e^{\lambda_0 t \w_l} -1\right) 
    \be\ul +{\bf \e^0} \sim  \frac{1}{r} e^{\lambda_0 t}  {\bf e}^{(0)}.
\enas
To estimate second moments, observe that,
given ${\cal F}_t$, $dX_0(t)$ is a centred Poisson innovation, so that
\bea\label{4+7}
\EE \left( d^2X_0(t)\right) = c \EE (M_{r-1}(t)) dt 
     \sim cr^{-1} e^{\lambda_0 t} \be\uo_{r-1}\,dt 
     = r^{-1}\l_0e^{\lambda_0 t}\,dt. 
\ena
But now, in~\Ref{1+5}, we can write
$$
M\uil(t) := \int_0^t(X_0(y)-1)\l_le^{\l_l(t-y)}\,dy 
   = \int_0^t\{e^{\l_l(t-v)} - 1\}\,dX_0(v),
$$
since $X_0$ is a.s.\ of bounded variation on finite intervals.  Hence,
from~\Ref{4+7},
\bea\label{4+77}
\EE |M\uil(t)|^2 
   \le 2r^{-1}\l_0\int_0^t e^{\l_0v}\{1+e^{2\Re\l_l(t-v)}\}\,dv.
\ena
For $1\le l\le r-1$, we can bound the integral in~\Ref{4+77}, uniformly
in $l$ and~$\rho$, in terms of~$\w_*$ and $\l_0(\rho)t$.  If $1\le r\le
4$, so that 
$\w_* = 0$, it is immediate that $\EE |M\uil(t)|^2 \le
4r^{-1}e^{\l_0t}$.
If $r=5$, so that $0<\w_*<1/2$, the bound has an extra factor of
$1/(1-2\w_*)$;
if $r=6$, then $\w_*=1/2$, and  $\EE |M\uil(t)|^2 \le
4r^{-1}\l_0te^{\l_0t}$;
for $r\ge 7$, $\EE |M\uil(t)|^2 \le
4r^{-1}(2\w_*-1)^{-1}e^{2\l_0\w_*t}$.
In all cases, for all $1\le l\le r-1$, we have 
$\EE |M\uil(t)|^2 = o(e^{2\l_0t})$, and since also 
$$
\EE |M^{(1,0)}(t)|^2 \le r^{-1}\l_0\int_0^t e^{2\l_0t-\l_0v}\,dv \le
r^{-1} e^{2\l_0t},
$$
it therefore follows easily that
\bea\label{4+9}
\EE M_0^2(t) = O(e^{2\l_0t})\quad \mbox{as }t \to \infty,
\ena
with the constant implied in the order estimate uniform for all~$\rho$.

The convergence of $\bW(t)$ can now be proved by second moment
arguments.
It follows directly from~\Ref{4+7} that $\var X_0(t) =
r^{-1}(e^{\lambda_0 t}
-1)$, and thus, using Kolmogorov's inequality on the martingale~$X_0$,
we
can easily show that, for any $\dd>0$,
\bea\label{4+6}
\sup_{t>0} |X_0(t)-1|e^{-(\l_0+\dd)t/2} < \infty.
\ena
Now, from~\Ref{1+5},
\bea
\bW(t) &=& e^{-\lambda_0 t}\bM(t)\non\\
&=& e^{-\lambda_0 t} \EE \bM(t) + e^{-\lambda_0 t}(X_0(t)-1) {\bf
\e^0}\non \\
&& +  
\int_0^t (X_0(y)-1) \sum_{l=0}^{r-1} \be\ul\mu_l \lambda_l
e^{\lambda_l(t-y)}
 dy e^{-\lambda_0 t},  \label{1+7}
\ena
and we consider the various terms in~\Ref{1+7}, using~\Ref{4+6}.
First, it is immediate that 
$$
\lim_{t\to\infty} e^{-\l_0t}(X_0(t)-1) = 0  \mbox{  a.s.}
$$
Then, distinguishing the cases $\Re(\l_l) \le 0$, $0 < \Re(\l_l) \le
\l_0/2$
and $\l_0/2 < \Re(\l_l) < \l_0$, it again follows from~\Ref{4+6} that,
for $1\le l\le r-1$,
$$
 \lim_{t\to\infty} \int_0^t(X_0(y)-1)\m_l\l_l e^{\l_l(t-y)-\l_0t}\,dy =
   0  \mbox{  a.s.}
$$
Finally, for $l=0$, we have
$$
 \lim_{t\to\infty} \int_t^\infty (X_0(y)-1)\l_0 e^{-\l_0y}\,dy = 0  
    \mbox{  a.s.}
$$
Hence, letting $t\to\infty$ in~\Ref{1+7}, it follows that the limit
\beas
\lim_{t\to\infty}\bW(t) 
 &=& {\bf e}^{(0)} \frac{1}{r} \left\{ 1 +  \int_0^\infty \lambda_0
(X_0(y)-1)
    e^{- \lambda_0 y} dy   \right\} = W\be\uo
\enas 
exists a.s.  To show that $\PP[W > 0] = 1$, note that
$$
W = \sum_{l\ge1} e^{-\l_0U_l} W\ul,
$$
where $(U_l,\,l\ge1)$ are the times of the births of new neighbourhoods
as children of the original neighbourhood around~$P$, and
$(W\ul,\,l\ge1)$
are independent copies of~$W$.  This immediately implies that $\PP[W=0]
\in\{0,1\}$, and $\PP[W=0]=1$ is excluded by $\EE W = 1/r$.

Note also that, again because~$X_0$ is locally of bounded variation,
\beas
\lefteqn{\bW(\infty) - \bW(t)}\\
&=& \mu_0 \be\uo e^{-\lambda_0 t} - \sum_{l=1}^{r-1}\be\ul \mu_l 
e^{-\lambda_0 t}
\left( e^{\lambda_l t} -1 \right)\\ 
&&- {\bf \e^0} e^{-\lambda_0 t}  
      + \mu_0 \be\uo  \int_t^{\infty} e^{-\lambda_0 y} dX_0(y) \\
&& \qquad- 
\sum_{l=1}^{r-1} \be\ul\mu_l \lambda_l e^{-(\lambda_0 - \lambda_l)t}
\int_0^t
\frac{ e^{-\lambda_l y} - e^{-\lambda_l t} }{\lambda_l} dX_0(y) \\
&=& \mu_0 \be\uo e^{-\lambda_0 t} - \sum_{l=1}^{r-1}\be\ul \mu_l 
        e^{-\lambda_0 t} \left( e^{\lambda_l t} -1 \right)\\
&&- {\bf \e^0} e^{-\lambda_0 t}  
      + \mu_0 \be\uo  \int_t^{\infty} e^{-\lambda_0 y} dX_0(y) \\
&& \qquad- \sum_{l=1}^{r-1} \be\ul\mu_l  \int_0^t dX_0(y) 
 \left\{e^{-\lambda_0 t + \lambda_l  (t-y)} - e^{-\lambda_0 t} 
\right\}.
\enas 
Thus
\bea\label{1+8}
(j!)^{-1}\l_0^j |\EE(W_j(\infty) - W_j(t))| \le c_{r1}
e^{-\l_0(1-\w_*)t}
\ena
and, from~\Ref{4+7},
\bea\label{1+9}
\var\{(j!)^{-1}\l_0^j (W_j(\infty) - W_j(t))\} \le
c_{r2}e^{-2\l_0(1-\w_*)t}
  \{1 + (\l_0t)1_{\{r=6\}}\},
\ena
for some constants $c_{r1}$ and~$c_{r2}$ and for all $0\le j\le r$,
completing the proof.
\ep
  
Now choose 
\bea\label{tauchoice}
\tau_x := {1\over \l_0(\rho)} \left\{ \frac12\log(L\rho) + x\right\},
\ena
for any~$x$ such that $|x| \le \frac14\log(L\rho)$, 
%agreeing with the previous choice 
much as in the case $r=1$.  Let $\whv_x$ as before be the
number of pairs of overlapping neighbourhoods, one from each of two 
independent pure growth processes $\bM(\tau_x)$ and~$\bN(\tau_x)$
and neither contained 
\vbox{
\noindent entirely in the other.
Then, from~\Ref{1+mean} and \Tr{Poisson}, it follows easily that
\bea
&&\left|\PP[\whv_x=0] - \EE\exp\left\{
   -L^{-1}\Bl\sum_{l=0}^r \a_l M_l(\tau_x) N_{r-l}(\tau_x)
	                                      \right.\right.\right.\non\\ 
&&\qquad\qquad\qquad\left.\left.\left.\mbox{}	
     - v(K)\int_0^{\t_x}(N_r(u)dM_0(u) + M_r(u)dN_0(u))\Br
	    \right\}\right|\non\\
&&\qquad = O((L\rho)^{-1/2}\log^r(L\rho)e^x), \label{vbnd}
\ena
}
since the probability of a given pair of neighbourhoods overlapping
cannot exceed
$$
L^{-1}\kappa v(K) \tau_x^r = O((L\rho)^{-1}\log^r(L\rho)),
$$
for some constant~$\kappa$ depending on the shape of~$K$, and because
$\EE M_0(\tau_x) \sim r^{-1}(L\rho)^{1/2}e^x$ from \Tr{r-D1}.
Then, with a proof much as for Lemma~\ref{cor3}, 
bounding~$M_0$ above by a Yule process with rate $\r v(K) r \t_x^{r-1}$
and noting that we now only have $\PP[A^l(i) = l] \le 1/l$, because in 
higher dimensions more recent neighbourhoods grow more slowly, it follows 
that
\bea\label{VVbnd}
\PP[\whv_x \neq V_x] = O((L\rho)^{-1/2}\log^{2r}(L\rho)e^{3x}),
\ena
where $V_x$ is the number of times that pairs of neighbourhoods
of the two associated growth and merge processes $R$ and~$R'$
meet before~$\t_x$.  These observations lead to the following theorem.

\begin{theorem}\label{r-D2}
Let $D$ denote the distance between two randomly chosen points of~$C$
on the graph with a $\Po(L\rho/2)$--distributed random number of
shortcuts.   Then
\beas
&&\PP\left[D > {2\over \l_0(\rho)} 
      \left\{ \frac12\log(L\rho) + x\right\}\right]\\
&&\qquad  =  \EE\left( \exp\left\{-e^{2x}WW'\left[{1\over v(K)}
    \sum_{l=0}^r{\a_l\over{r\choose l}} - 1\right]\right\}\right) \\
&&\qquad\qquad  
 + O\left\{(L\rho)^{-(1-\w_*)/2}\log^{2r+1}(L\rho)(e^{3x}+e^x)\right\},
\enas
uniformly in $|x| \le \frac14\log(L\rho)$, where $W$ and~$W'$ are
independent copies of the limiting random variable of \Tr{r-D1}.
\end{theorem}

\proof   In view of \Ref{vbnd} and~\Ref{VVbnd}, it is enough to show
that
\beas
\lefteqn{\EE\exp\left\{-L^{-1}\Bl\sum_{l=0}^r \a_l M_l(\tau_x)N_{r-l}(\t_x) 
   - v(K)\int_0^{\t_x}(N_r(u)dM_0(u) + M_r(u)dN_0(u)) \Br\right\} }\\
  &&
  = \EE\left( \exp\left\{-e^{2x}WW'\left[{ 1\over v(K)}
         \sum_{l=0}^r{\a_l\over{r\choose l}} - 1\right]\right\}\right) 
  + O(\l_0\tau_x\exp\{2x-\l_0(1-\w_*)\tau_x\}).
\enas
However, direct calculation shows that
$$
L^{-1}\sum_{l=0}^r \a_l M_l(\tau_x) N_{r-l}(\tau_x)
  = {e^{2x} \over r!v(K)} \sum_{l=0}^r \a_l \{\l_0^l W_l(\tau_x)\}
   \,\{\l_0^{r-l}W'_{r-l}(\tau_x)\}
$$
and that
\beas
\lefteqn{L^{-1}v(K)\int_0^{\t_x}(N_r(u)dM_0(u) + M_r(u)dN_0(u))}\\
&=& L^{-1}v(K)\Blb \vphantom{\int_0^{\t_x}}
   e^{2\l_0\t_x}\{W_0(\t_x)W_r'(\t_x) 
          + W_0'(\t_x)W_r(\t_x)\right.\\
&&\qquad \left. \mbox{}- r\int_0^{\t_x} e^{2\l_0u}\{W_0(u)W'_{r-1}(u)
    + W_0'(u)W_{r-1}(u)\}\,du \Brb\\
&=& e^{2x}\Bl W_0(\t_x)\{\r v(K)W_r'(\t_x)\} 
    \vphantom{I_1^1}+ W'_0(\t_x)\{\r v(K)W_r(\t_x)\}
    \Br\\
&&\qquad \mbox{} - \frac1{L\r}\int_0^{\t_x} \l_0e^{2\l_0u}
		   \Bl W_0(u)\{\l_0^{-1}r\r v(K)W'_{r-1}(u)\}\right. \\
&&\qquad\qquad\qquad\qquad\qquad\qquad\left.\mbox{}			 
     + W'_0(u)\{\l_0^{-1}r\r v(K)W_{r-1}(u)\} \Br\,du 
\enas
where, by \Tr{r-D1},
$$
\EE|(j!)^{-1}\l_0^jW_j(t)-W| \le c_r(1+(\l_0t)^{1/2}1_{\{r=6\}})
  e^{-\l_0(1-\w_*)t},\quad 0\le j\le r,
$$
in which, for $j = 0,r-1$ and~$r$, the left hand side takes
the form
$\EE|W_0(t)-W|$, $\EE|\l_0^{-1}
  r\r v(K) W_{r-1}(t) - W|$ and $\EE|\r v(K)W_r(t) - W|$ respectively.
Then the inequality $|e^{-a} - e^{-b}| \le |a-b|$ for all $a,b\ge0$
completes the proof.
\ep

Note that in the $r$--dimensional model, the distances~$D$ have 
$\l_0(\rho)$ in the denominator, of order $O(\rho^{1/r})$,
in place of~$2\rho$ in the case
$r=1$.  This scaling can be understood as follows.  If there were
no shortcuts, the average distance between pairs of points would be
of order~$L^{1/r}$.  With about $L\rho/2$ shortcuts, this distance
is reduced by a factor of order $(L\rho)^{-1/r}\log(L\rho)$.  Thus,
in a higher dimensional space, the reduction in distance as a result
of introducing shortcuts is correspondingly smaller.

\bigskip 
{\bf Acknowledgement.} The authors would like to thank A.G. Pakes
for sharing his expertise on the limiting variable $W$. Moreover
the authors are grateful to the referees for an extremely careful
reading of the paper, and for many valuable comments.

\vfil\eject

\end{document}